\documentclass[12pt,]{nature}
\usepackage[english]{babel}
\usepackage{tabulary,graphicx,amsmath,amsfonts,amssymb}
\usepackage[utf8x]{inputenc}
\usepackage{tikz}
\usepackage{braket}
\usepackage{pgfplots}
\usepackage{csquotes}
\usepackage{hhline}

\usepackage{amsthm}

\usepackage{dcolumn}

\usepackage{tabularx}
\usepackage[colorlinks=true,linkcolor=blue,citecolor=blue,urlcolor=blue]{hyperref}
\usepackage{listings}
\usepackage{longtable}
\usepackage{float}
\usepackage{epsfig}
\usepackage{subcaption}
\makeatletter
\renewenvironment{figure}
               {\@float{figure}}
               {\end@float}
\renewenvironment{figure*}
               {\@dblfloat{figure}}
               {\end@dblfloat}
\renewenvironment{table}
               {\@float{table}}
               {\end@float}
\renewenvironment{table*}
               {\@dblfloat{table}}
               {\end@dblfloat}
\makeatother

\usepackage{url,multirow,morefloats,floatflt,cancel,tfrupee}
\makeatletter

\AtBeginDocument{\@ifpackageloaded{textcomp}{}{\usepackage{textcomp}}}
\makeatother
\usepackage{colortbl}
\usepackage{xcolor}
\usepackage{pifont}
\usepackage[nointegrals]{wasysym}
\makeatletter

\def\mcWidth#1{\csname TY@F#1\endcsname+\tabcolsep}

\def\cAlignHack{\rightskip\@flushglue\leftskip\@flushglue\parindent\z@\parfillskip\z@skip}
\def\rAlignHack{\rightskip\z@skip\leftskip\@flushglue \parindent\z@\parfillskip\z@skip}

\if@twocolumn\usepackage{dblfloatfix}\fi 
\AtBeginDocument{
\expandafter\ifx\csname eqalign\endcsname\relax
\def\eqalign#1{\null\vcenter{\def\\{\cr}\openup\jot\m@th
  \ialign{\strut$\displaystyle{##}$\hfil&$\displaystyle{{}##}$\hfil
      \crcr#1\crcr}}\,}
\fi
}

\let\lt=<
\let\gt=>
\def\processVert{\ifmmode|\else\textbar\fi}

\@ifundefined{subparagraph}{
\def\subparagraph{\@startsection{paragraph}{5}{2\parindent}{0ex plus 0.1ex minus 0.1ex}%
{0ex}{\normalfont\small\itshape}}%
}{}

\newcommand\role[1]{\unskip}
\newcommand\aucollab[1]{\unskip}
  
\@ifundefined{tsGraphicsScaleX}{\gdef\tsGraphicsScaleX{1}}{}
\@ifundefined{tsGraphicsScaleY}{\gdef\tsGraphicsScaleY{.9}}{}
\def\checkGraphicsWidth{\ifdim\Gin@nat@width>\linewidth
    \tsGraphicsScaleX\linewidth\else\Gin@nat@width\fi}

\def\checkGraphicsHeight{\ifdim\Gin@nat@height>.9\textheight
    \tsGraphicsScaleY\textheight\else\Gin@nat@height\fi}

\def\fixFloatSize#1{}%\@ifundefined{processdelayedfloats}{\setbox0=\hbox{\includegraphics{#1}}\ifnum\wd0<\columnwidth\relax\renewenvironment{figure*}{\begin{figure}}{\end{figure}}\fi}{}}
\let\ts@includegraphics\includegraphics

\def\inlinegraphic[#1]#2{{\edef\@tempa{#1}\edef\baseline@shift{\ifx\@tempa\@empty0\else#1\fi}\edef\tempZ{\the\numexpr(\numexpr(\baseline@shift*\f@size/100))}\protect\raisebox{\tempZ pt}{\ts@includegraphics{#2}}}}

\AtBeginDocument{\def\includegraphics{\@ifnextchar[{\ts@includegraphics}{\ts@includegraphics[width=\checkGraphicsWidth,height=\checkGraphicsHeight,keepaspectratio]}}}

\def\URL#1#2{\@ifundefined{href}{#2}{\href{#1}{#2}}}

\def\UrlOrds{\do\*\do\-\do\~\do\'\do\"\do\-}%
\g@addto@macro{\UrlBreaks}{\UrlOrds}
\makeatother

\def\fixFloatSize#1{}

\newcolumntype{C}{>{\centering\arraybackslash}X}
\pgfplotsset{compat=1.14}
\begin{document}
\setcounter{secnumdepth}{3}
\title{Demonstration of teleportation-based error correction in the IBM quantum computer}

\author{K.~M. Anandu$^{1}$\thanks{E-mail: anandu.k.madhu@gmail.com}{ },
              Muhammad Shaharukh$^{2}$\thanks{E-mail: sharkaj786@gmail.com}{ },
              Bikash K. Behera$^{4}$\thanks{E-mail: bkb18rs025@iiserkol.ac.in} {} \&
              Prasanta K. Panigrahi$^{4}$\thanks{Corresponding author.}\ \thanks{E-mail: pprasanta@iiserkol.ac.in}
    }
\maketitle 

\begin{affiliations}
  \item 
    Department of Physics\unskip, Cochin University of Science and Technology 
    \unskip, Kochi\unskip, 682022\unskip, Kerala\unskip, India
  \item 
    Department of Physics\unskip, Aligarh Muslim University\unskip, Aligarh\unskip, 202002\unskip, Uttar Pradesh\unskip, India
   \item 
    Department of Physical Sciences, Indian Institute of Science Education and Research Kolkata, Mohanpur, 741246, West Bengal, India
\end{affiliations}

\begin{abstract}
Quantum error correcting codes (QECC) are the key ingredients both for fault-tolerant quantum computation and quantum communication. Teleportation-based error correction (TEC) helps in detecting and correcting operational and erasure errors by performing X and Z measurements during teleportation. Here we demonstrate the TEC protocol for the detection and correction of a single bit-flip error by proposing a new quantum circuit. A single phase-flip error can also be detected and corrected using the above protocol. For the first time, we illustrate detection and correction of erasure error in the superconducting qubit-based IBM's 14-qubit quantum computer. 
\end{abstract}  


\section{Introduction}

Successful transmission of quantum information requires the information to be less affected by noise in the communication channel. Quantum error correction (QEC) \cite{qtec_GottesmanPRA1996,qtec_ChiaveriniNature2004,qtec_SingharXiv:1807.02883,qtec_CorcolesNatCommun2015} achieves this by encoding the information into a large number of physical qubits and by performing operations on them to detect and correct errors. QEC is an essential ingredient in achieving fault-tolerant quantum computation \cite{qtec_PreskillarXiv:9712048, qtec_GottesmanPRA1998,qtec_KnillPRSA1998,qtec_AharonovSJC2008}. Many protocols have been proposed such as the Shor code \cite{qtec_ShorPRA1995,qtec_VandersypenNature2001} and Steane code \cite{qtec_SteanePRL1996,qtec_SteanePRA2003} for error correction mechanism. Errors can be broadly divided into two types: (a) operational errors and (b) erasure errors \cite{qtec_ClevePRL1999}. Operational errors are the action of Pauli operators on qubits. Examples of operational errors are: (i) bit-flip error \cite{qtec_RisteNatCommun2015} (ii) phase-flip error \cite{qtec_ShengPRA2010} and (iii) combination of both bit-flip and phase-flip errors. Operational errors can be corrected by Quantum Error Correction Codes (QECC) \cite{qtec_CalderbankarXiv:9608006} such as the bit-flip code, the phase-flip code, the Shor code etc. Erasure errors are errors where photons or qubits (whose exact location information is known) are erased. A quantum erasure channel replaces a qubit (qudit) with an `erasure state' that is orthogonal to all the basis states of a qubit (qudit) with a certain probability, thereby erasing a qubit (qudit) and enabling the receiver know that it has been erased. Erasure error occurs physically due to various situations such as the leakage to other states, atom losses \cite{qtec_TheisPRL2004}, and photon losses \cite{qtec_DuanPRL2004}. By reliably measuring the logical operators of a code, we can actively detect and correct errors. Teleportation-based error correction (TEC) \cite{qtec_KnillNature2005} provides a method to do so. Bell measurements \cite{qtec_BoschiPRL1998,qtec_BarencoPRL1995,qtec_LutkenhausPRA1999} implemented during teleportation acts as syndrome measurements \cite{qtec_SteaneNature1999} here. Erasure errors can be corrected effectively by implementing TEC.

Protecting quantum information from erasure errors remains an open challenge in a long distance quantum communication as well as in practical quantum computation. Recently Muralidharan \emph{et al.} \cite{qtec_MuralidharanPRL2014} investigated the usage of highly effective error correcting codes of multilevel systems to protect encoded quantum information from erasure errors and then they implemented it to repetitively correct these errors. For successful long distance quantum communication through optical fibres \cite{qtec_GisinNatPhoton2007}, photon loss errors possess a major threat. They proposed three generations of quantum repeaters \cite{qtec_DuanPRL2004,qtec_BriegelPRL1998} based on different approaches to correct both photon loss and operational errors. One of them employs QEC to correct both loss and operational errors and it does not require a two-way classical communication between repeater stations, which provides a significant advantage over other protocols allowing for ultrafast communication across transcontinental distances \cite{qtec_EwertPRL2016,qtec_NamikiPRA2016,qtec_FowlerPRL2010,qtec_GlaudellNJP2016}. 

We explicate the teleportation-based error correction for bit-flip errors and erasure errors in IBM's 14-qubit quantum computer. We measure the fidelity of the states obtained for TEC of erasure errors using quantum state tomography \cite{qtec_ThewPRA2002,qtec_DavidPRL2010,qtec_VishnuQIP2018}. A number of experiments related to quantum information science have been performed by researchers on IBM quantum computer since its inception. Some of the works include quantum simulation \cite{qtec_LiertaarXiv2018,qtec_ZhukovQIP2018,qtec_KapilarXiv:1807.00521,qtec_ViyuelanpjQI2018,qtec_HegadearXiv:1712.07326}, quantum artificial intelligence \cite{qtec_RodriguezarXiv:1711.09442}, quantum machine learning \cite{qtec_ZhaoarXiv:1806.11463}, quantum teleportation \cite{qtec_FedortchenkoarXiv:1607.02398,qtec_SisodiaQIP2018,qtec_VishnuQIP2018}, quantum state discrimination \cite{qtec_MajumderarXiv:1803.06311,qtec_SatyajitQIP2018}, quantum error correction \cite{qtec_SingharXiv:1807.02883,qtec_HarperarXiv:1806.02359,qtec_WillscharXiv:1805.05227,qtec_GhoshQIP2018}, quantum algorithms \cite{qtec_GangopadhyayQIP2018,qtec_ColesarXiv:1804.03719}, quantum games \cite{qtec_PalRG.2.2.19777.86885,qtec_MahantiRG.2.2.28795.62241}, quantum circuit optimization \cite{qtec_ZhangarXiv:1807.01703}, quantum cryptography \cite{qtec_BeheraQIP2017} to name a few. It is made available to the public via the cloud through the platform IBM Q Experience \cite{qtec_IBM}. For implementing the circuit on the quantum computer, we use the Quantum Information Science Kit (QISKit) \cite{qtec_QISKit} provided by the IBM Quantum Experience platform.

\section{Results \label{qtec_Results}} 

\textbf{Implementation of TEC for operational errors}

Operational errors are errors which are caused by the action of Pauli operators on the encoded information \cite{qtec_GottesmanPRA2001,qtec_MilioneOptLett2015}. The action of Pauli-X ($\sigma_{x}$) \cite{qtec_BaconPRA2006,qtec_HeilmannSciRep2014} and Pauli-Z ($\sigma_{z}$) \cite{qtec_TuckettPRL2018} operators are called bit-flip and phase-flip errors respectively. The action of Pauli-Y operator is a combination of action of both the above errors. Let's represent $\sigma_{x}$, $\sigma_{y}$ and $\sigma_{z}$ operators as X, Y and Z respectively. Since Y can be represented as $Y=iXZ$, sequential correction of X error and Z error will automatically correct the Y error as well. Error correction of X and Z errors are done by identifying the type of error and the qubit location in which the error has occurred \cite{qtec_ChapmanNatCommun2016,qtec_AhsanJETC2016} and applying the corresponding correction operation (e.g., if an X error has occurred, the effect of the error can be nullified by applying an X operator to the same qubit, since $X^2=I$). The correction of these errors is a major challenge in the field of quantum computation as well as quantum communication. QECC helps in the detection and correction of such errors. We could actively detect and correct the errors by measuring the logical operators of the code reliably \cite{qtec_YoshidaPRA2010}. Teleportation-based error correction provides a method to do so.

\begin{figure}[]
\centering
\includegraphics[scale=1.0]{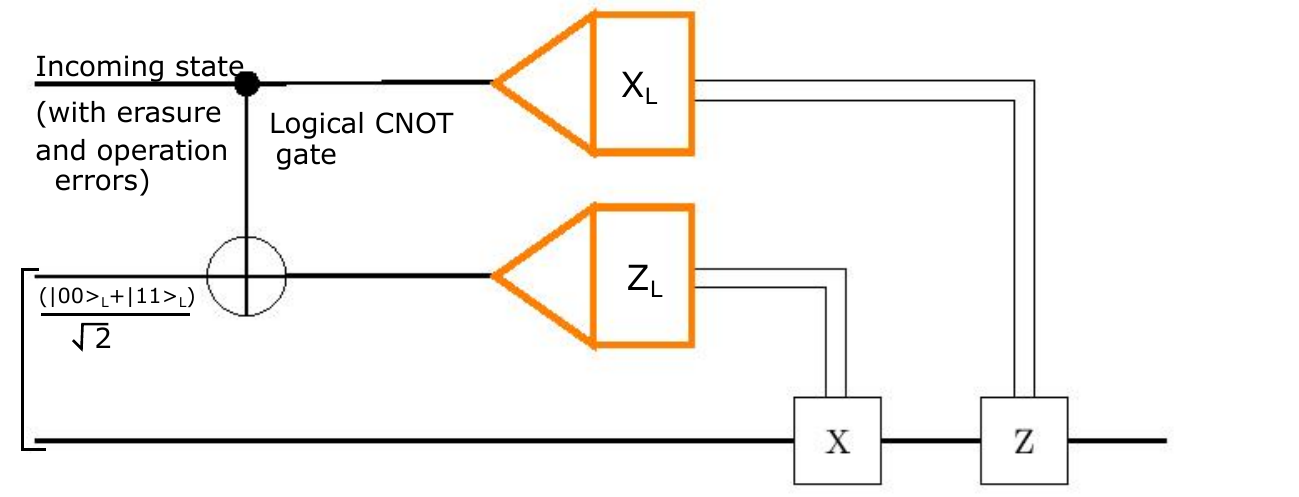} 
\caption{\textbf{Quantum circuit to implement teleportation-based error correction.} The first qubit contains the incoming state. A logical Bell state is prepared using the second and third qubit. $X_L$ and $Z_L$ denote the logical Bell measurements in X and Z bases respectively. The double line after the Bell measurement represents a classical communication channel. The X and Z boxes at the end of the third line represent the corresponding corrections made according to the outcome of the Bell measurements.}
\label{qtec_Fig1}
\end{figure}

The quantum circuit representation of TEC is illustrated in Fig. \ref{qtec_Fig1}. Here, $X_L$ and $Z_L$ are the logical Bell measurements \cite{qtec_ShiOptCommun2010}, which are employed during teleportation. They are used in detecting the errors that affect the encoded information. The X-basis (Z-basis) measurements determine whether a phase-flip (bit-flip) error has occurred or not. During teleportation, logical CNOT gates \cite{qtec_DengJOSAB2007} are performed between the incoming message and the first logical qubit of the Bell state. The error correction is applied to the second logical qubit of the Bell state as the information is transferred to that logical qubit during teleportation. Classical control operations are implemented during teleportation which are shown by X and Z boxes in Fig. \ref{qtec_Fig1}.

In the proposed TEC circuit, a logical qubit is encoded in three physical qubits. Our QECC can correct one bit-flip error on an arbitrary qubit. It requires a total of eleven qubits. The quantum circuit for TEC of operational errors is illustrated in Fig. \ref{qtec_Fig2}. The detailed step-by-step procedure is given as follows. 

\begin{enumerate}

\item We encode a logical qubit, $\alpha \ket{0}_L+\beta \ket{1}_L$ as

\begin{equation} \label{qtec_Eq1}
\alpha \ket{0}_L+\beta \ket{1}_L \rightarrow \alpha \ket{000} +\beta \ket{111}
\end{equation}

where $\alpha=0.92$ and $\beta=0.38$ as depicted in block A of the Fig. \ref{qtec_Fig2}.

\item The logical Bell state is encoded as

\begin{equation} \label{qtec_Eq2}
(\Ket{00}_L+\Ket{11}_L)/\sqrt{2}\rightarrow (\Ket{000000}+\Ket{111111})/\sqrt{2}
\end{equation}

The quantum circuit for this encoding is illustrated in block B of Fig. \ref{qtec_Fig2}.

\item As illustrated in the block C of the Fig. \ref{qtec_Fig2}, logical CNOT gates are performed between the encoded information and the first logical qubit of the Bell state.

\item The Bell-basis measurement of the logical operators of the code for the error detection and the corresponding error correction using unitary gates could not be implemented here due to the limitation of the architecture of the device. Instead we employ stabilizer measurements \cite{qtec_chuang2010} for the detection and correction of errors which are performed non-destructively using two ancilla qubits and they provide the error syndrome \cite{qtec_ReedNature2012,qtec_TamakiPRA2010} for error correction. This is achieved by performing a `collective measurement' on the first and second qubits and on the first and third qubits of the first logical qubit of the Bell state simultaneously. This is implemented using four CNOT gates between the first logical qubit of the Bell state and the ancilla qubits as demonstrated in the block D of the Fig. \ref{qtec_Fig2}. These are stabilizer measurements \cite{qtec_RisteNatCommun2015} of $Z_1Z_2$ and $Z_1Z_3$ respectively. We could understand on which qubit the error has occurred by looking at the stabilizer measurements \cite{qtec_RisteNatCommun2015}. 

\item Based on these measurement results, error correction operations are applied on the first logical qubit of the Bell state using CNOT and Toffoli gates \cite{qtec_LinPRA2009}. This is illustrated in the block D of the Fig. \ref{qtec_Fig2}. If there is a bit-flip error on the first qubit, then both the ancilla qubits are flipped to the $\Ket{1}$ state. This error is corrected by flipping the first qubit using a Toffoli gate conditioned on the ancilla qubits and targeted on the first qubit. Due to the action of the first two CNOT gates which are targeted on the second and third qubits and conditioned on the ancilla qubits, the second and third qubits are also flipped. The first two Toffoli gates which are targeted on the second and third qubit prevent the respective qubits from flipping. If the bit-flip has occurred on the second qubit, then the first ancilla qubit is in the $\Ket{1}$ state and the second ancilla is in the $\Ket{0}$ state. Therefore, the first CNOT gate conditioned on the first ancilla flips the second qubit and the first Toffoli gate conditioned on these ancilla qubits remains ineffective. Similarly, if error happens on the third qubit, only the second ancilla is in the $\Ket{1}$ state. Therefore, the second CNOT gate flips it to the original state and the second Toffoli gate conditioned on these ancilla qubits remains ineffective. 

\item The quantum conditional operators corresponding to the teleportation is demonstrated in the block F of the Fig. \ref{qtec_Fig2}.

\item The measurement devices at the end of the circuit represents the measurement of the second logical qubit of the Bell state in the Z-basis, which gives back the encoded information.

\begin{figure}[]
\centering
\includegraphics[scale=0.335]{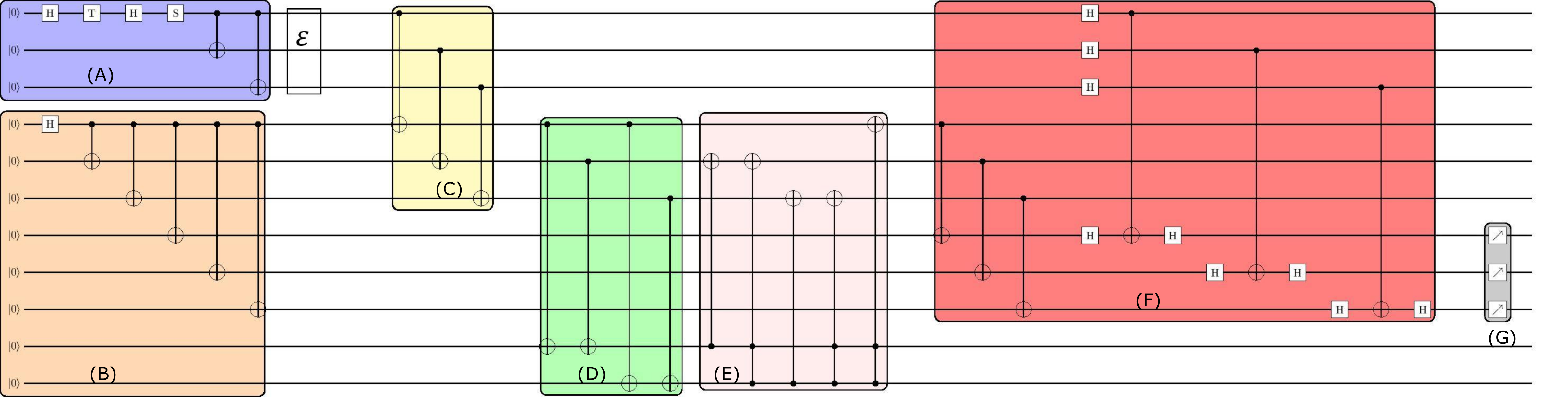}
\caption{\textbf{Quantum circuit to implement TEC of bit-flip error.} The encoding of the information is done using the circuit in block A (blue). The logical Bell state is prepared using the circuit in block B (orange). The box named `$\varepsilon$'  between the blocks A and C represents an error channel where a single bit-flip error affects an arbitrary qubit. The stabilizer measurements which are used for the detection and correction of errors are demonstrated in the block D (green). The error correction performed by CNOT and Toffoli gates using stabilizer measurements are illustrated in block E (pink). The quantum conditional operators are shown in the block F (red). The measurement devices at the end of the circuit in block G represents measurements done on the second logical qubit of the Bell state in the Z-basis.}
\label{qtec_Fig2}
\end{figure}
The same circuit can be used for TEC of phase-flip errors by changing the basis of encoding from Z-basis to X-basis. This can be achieved by placing Hadamard gates before and after the box named `$\varepsilon$' where qubits get affected by errors.
\end{enumerate}

\textbf{Implementation of TEC for erasure errors}

Erasure errors \cite{qtec_ClevePRL1999} are defined as errors which occur due to the `erasure' of a qubit, whose location information is known. Examples of erasure errors are photon loss errors or a two-level atomic system \cite{qtec_HetetPRL2008} where unwanted levels are coupled to the `allowed' two-levels of the system. Erasure errors can also occur during the transmission of a message through a quantum erasure channel during quantum communication. Erasure errors can be corrected by implementing TEC. Logical CNOT gates are implemented between the incoming message with erasure and operational errors and the first logical qubit of the Bell state. The quantum nondemolition (QND) \cite{qtec_SewellNatPhoton2013,qtec_LupascuNatPhys2007} measurement results give us the location information of the qubits that was erased. The Bell measurements employed during teleportation acts as logical operations on the encoded bits which provides the error syndrome of the code. Based on the Bell measurement results, we adjust the Pauli frame \cite{qtec_KnillNature2005} and the syndrome information obtained from Bell measurements is used to apply corrective unitary operations on the second logical qubit of the Bell state which contains the outgoing message. 

\begin{figure}[]
\includegraphics[scale=0.183]{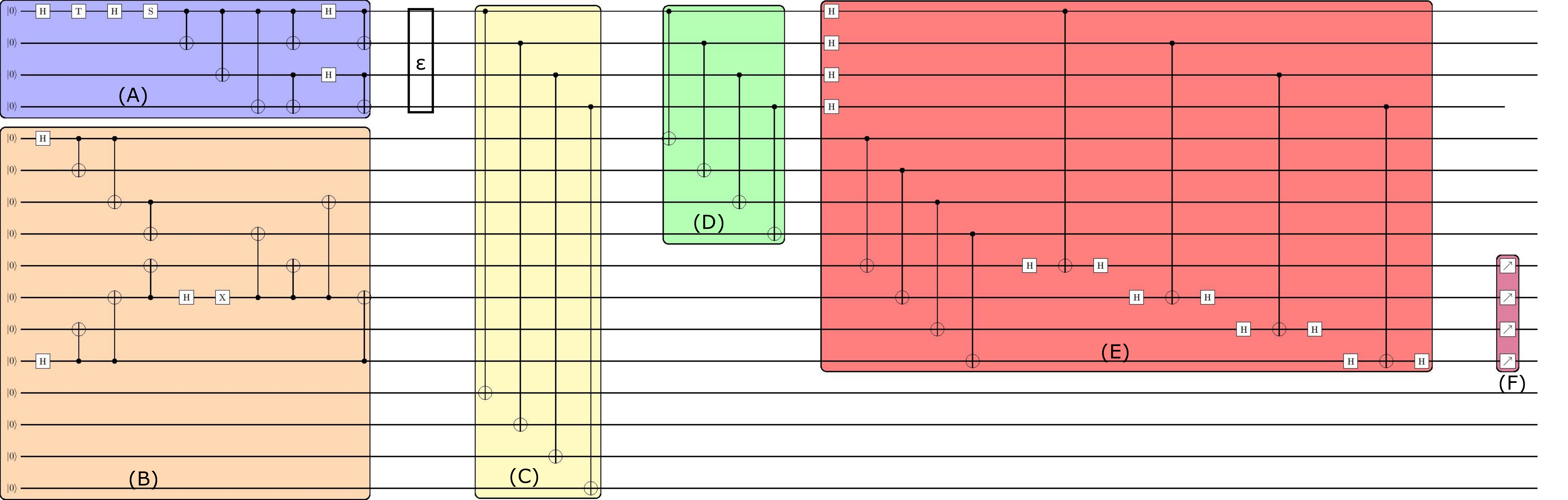}
\caption{\textbf{Quantum circuit for TEC of erasure errors.} The encoding of the information is performed using circuit in block A (blue). The logical Bell state is prepared using circuit illustrated in block B (orange). The box labelled `$\varepsilon$' indicates an erasure channel where out of the four encoded qubits, an arbitrary qubit is erased. QND measurement of incoming message is given in block C (yellow). The logical CNOT gates implemented during teleportation is given in block D (green). The quantum conditional operators are implemented in the circuit as illustrated in block E (red). The measurement devices in block F (pink) at the end of the circuit measures the logical operators in the Z-basis.}
\label{qtec_Fig3}
\end{figure}

In the proposed TEC circuit for erasure error, we encode the message based on redundancy and parity encoding \cite{qtec_RalphPRL2005,qtec_RubensteinComputNetw2004}. Our QECC can correct one erasure error on an arbitrary qubit. The total number of qubits required is sixteen. The quantum circuit for TEC of erasure error is given in Fig. \ref{qtec_Fig3}. The step-by-step procedure for erasure error correction is provided below. 

\begin{enumerate}
\item We encode a logical qubit, $\alpha \ket{0}_L+ \beta \ket{1}_L$ as

\begin{equation} \label{qtec_Eq3}
    \alpha \ket{0}_L+ \beta \ket{0}_L \rightarrow \frac{\alpha}{2} (\ket{00}+\Ket{11})_{12}(\ket{00}+\Ket{11})_{34} + \frac{\beta}{2} (\ket{00}-\Ket{11})_{12}(\ket{00}-\Ket{11})_{34}
\end{equation}
where $\alpha=0.92$ and $\beta=0.38$. The circuit for encoding is demonstrated in block A of the Fig. \ref{qtec_Fig3}.

\item The logical Bell state is encoded as

\begin{eqnarray} \label{qtec_Eq4}
  (\Ket{00}_L+\Ket{11}_L)/\sqrt{2} &\rightarrow    \frac{1}{4\sqrt{2}}[(\ket{00}+\Ket{11})_{12}(\ket{00}+\Ket{11})_{34} (\ket{00}+\Ket{11})_{56}(\ket{00}+\Ket{11})_{78} \nonumber \\ & + (\ket{00}-\Ket{11})_{12}(\ket{00}-\Ket{11})_{34}(\ket{00}-\Ket{11})_{56}(\ket{00}-\Ket{11})_{78}]  
\end{eqnarray}

which is illustrated in block B of the Fig. \ref{qtec_Fig3}.

\item Quantum nondemolition (QND) measurements are performed using CNOT gates conditioned on the incoming message and targeted on ancilla qubits to obtain the location information of the erased qubits. This is identified as such when a qubit is erased, the conditional not is not applied between the corresponding information qubit and the ancilla qubit \cite{qtec_NamikiPRA2016}. This circuit is demonstrated on block C of the Fig. \ref{qtec_Fig3}. 

\item The logical CNOT gates are illustrated in the block D of the Fig. \ref{qtec_Fig3}, where they are implemented in between the incoming message and the first logical qubit of the Bell state.

\item Since our QEC corrects only a single erasure error on the encoded message, we implement the Bell measurements and the corresponding unitary operators are operated on the second logical qubit of the Bell state as quantum conditional operators \cite{qtec_chuang2010} as illustrated in block E of the Fig. \ref{qtec_Fig3}. 

\item In our circuit, we assume that an erasure channel acts in between the blocks A and C. This is illustrated by a block labelled `$\epsilon$' in Fig. \ref{qtec_Fig3}. The simulation of erasure of an error can be achieved by removing the gates which affect the qubit that need to be erased.

\item The X logical operators and the Z logical operators of our code are $X_L=IZIZ,ZIZI$ and $Z_L=IIXX,XXII$ respectively. We will get back the encoded message by measuring the logical operators of our code reliably. For example, if the first or third qubit is lost, we measure the logical operators $X_L=IZIZ$. And if the second or fourth qubit is lost, we measure the logical operators $X_L=ZIZI$.

\end{enumerate}

\begin{figure}[]
\centering
\includegraphics[scale=0.5]{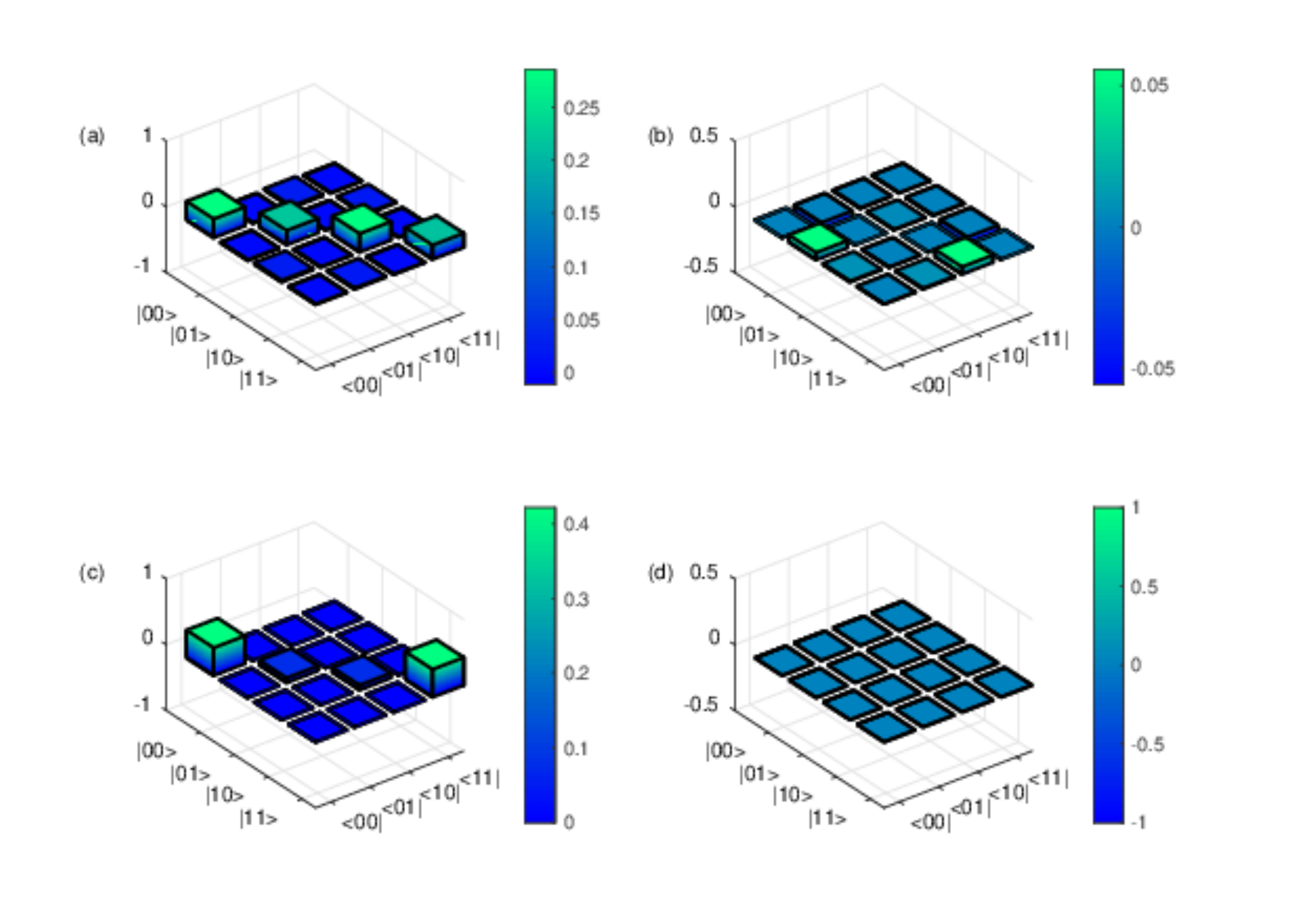} 
\caption{The figure represents the real and imaginary parts of experimental and ideal density matrices for the implementation of TEC of erasure errors. (a),(b): Experimental case; (c),(d): Ideal case. }
\label{qtec_Fig4}
\end{figure}

The theoretical and experimental density matrices of the error corrected state are depicted in Fig. \ref{qtec_Fig4}. The accuracy of the output state can be understood by comparing the theoretical and experimental density matrices. This can be achieved by calculating the fidelity between the ideal state and the experimentally prepared state. The fidelity of the experimental result is 0.8325.

\section{Discussion \label{qtec_Discussion}}
We have successfully implemented the Teleportation based error correction (TEC) for operational errors and erasure errors on the IBM's 14-qubit quantum computer. We have also quantified the fidelity of the output states generated for TEC of erasure errors using quantum state tomography. The TEC protocol for operational errors can be used in correcting a single bit-flip error acting on an arbitrary qubit of the encoded state and the TEC for erasure error can be used for correcting an erasure error acting on an arbitrary qubit of the encoded state. Parity and redundancy encoding is used for encoding the information in the case of erasure errors. Erasure errors affecting the transmission of information over long distances can be corrected using the TEC protocol that we proposed. The TEC based protocols in general can be implemented in any system for the detection and correction of operational and erasure errors affecting the system. In future, this work can be extended to quantify how gate errors and memory decoherence affect the error correction for erasure errors. 

\section{Methods\label{qtec_Methods}}

%\textbf{Simulation of the protocol on the 14-qubit IBM quantum computer.}
\begin{figure}[]
\centering
\includegraphics[width=0.675\linewidth]{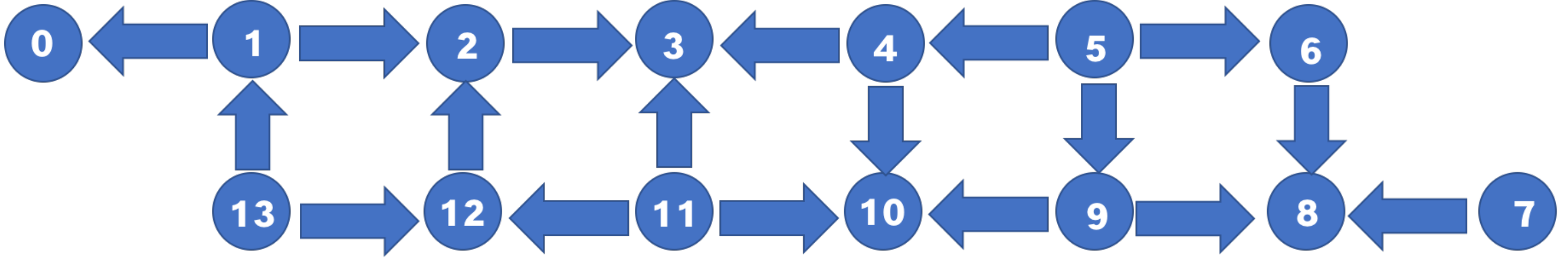}
\caption{\textbf{Architecture of IBM's 14 Q quantum processor.} The figure shown here is the layout of the 14-qubit quantum processor Melbourne [ibmq$\_$16$\_$melbourne]. The 14 qubits are connected through $CNOT$ operations and the allowed connection of $CNOT$ are $Q1\longrightarrow [Q0,Q2]$, $Q2\longrightarrow Q3$, $Q4\longrightarrow Q10$, $Q5\longrightarrow [Q9,Q6,Q4]$, $Q6\longrightarrow Q8$, $Q7\longrightarrow Q8$, $Q9\longrightarrow [Q8,Q10]$, $Q11\longrightarrow [Q12,Q10,Q3]$, $Q12\longrightarrow [Q2]$, $Q13\longrightarrow [Q12,Q1]$, where $Qi\longrightarrow Qj$ denotes $Qi$ and $Qj$ are the control and target qubit respectively.}
\label{qtec_Fig5}
\end{figure}

\begin{table}[H]
\centering
\begin{tabular}{c c c}
\hline
\hline
Qubits & $\omega^{R}_i\dagger/2\pi$ (GHz) & $\omega_i\ddagger/2\pi$ ($GHz$) \\
\hline
Q0 & 6.95518 & 5.1000 \\ 
Q1 & 7.05693 & 5.2384 \\ 
Q2 & 6.97179 & 5.0328 \\
Q3 & 7.04784 & 4.8961 \\
Q4 & 6.94523 & 5.0262 \\
Q5 & 7.07587 & 5.0670 \\
Q6 & 6.95297 & 4.9237 \\
Q7 & 6.96377 & 4.9744 \\
Q8 & 7.04930 & 4.7381 \\
Q9 & 6.96707 & 4.9633 \\
Q10 & 7.05513 & 4.9450 \\
Q11 & 6.95492 & 5.0046 \\
Q12 & 7.06722 & 4.7598 \\
Q13 & 6.94433 & 4.9685 \\ 
\hline
\hline
\end{tabular}\\

$\dagger$ Resonance Frequency of the Readout Resonator, $\ddagger$ Qubit Frequency.
\caption{\textbf{Specifications of the parameters of each qubit in the `ibmq$\_$16$\_$melbourne' quantum computer}}
\label{qtec_Tab1}
\end{table} 

\textbf{TEC of operational errors} 
We encode a data qubit as given in Eq. \eqref{qtec_Eq1}. The encoding is illustrated in the block A of Fig. \ref{qtec_Fig2}. 
To encode the information $\alpha=0.92$ and $\beta=0.38$, which can be represented as $\alpha=\cos{\frac{\pi}{8}}$ and $\beta=\sin{\frac{\pi}{8}}$, we employ three gates, $T_0$, $H_0$, $S_0$ in succession. It is to be noted that $A_i$ is a single qubit gate where the operation A acts on the $i$th qubit and $CNOT_{ij}$ is the controlled not gate where i and j are the control and target qubits respectively. The $H_0$ gate is used to create a superposition of $(\Ket{0}+\Ket{1})/\sqrt{2}$ from the initial $\Ket{0}$ state. The $CNOT_{0,1}$ and $CNOT_{0,2}$ gates of the block are used to encode a single qubit $(\Ket{0}+\Ket{1})/\sqrt{2}$ into a logical qubit of the form $(\Ket{000}+\Ket{111})/\sqrt{2}$. This circuit is used for encoding the information in the form of physical qubits in our TEC protocol. The Bell state is encoded as given in Eq. \eqref{qtec_Eq2}. The quantum circuit for this encoding is illustrated in block B of Fig. \ref{qtec_Fig2}. The first $H_3$ gate is used to create a superposition of $(\Ket{0}+\Ket{1})/\sqrt{2}$ from the initial $\Ket{0}$ state. The  $CNOT_{3,4}$, $CNOT_{3,5}$, $CNOT_{3,6}$, $CNOT_{3,7}$ and $CNOT_{3,8}$ gates are used to create a logical state of the form $(\Ket{000000}+\Ket{111111})/\sqrt{2}$ from the state $(\Ket{0}+\Ket{1})/\sqrt{2}$. We use this circuit to create a logical Bell state in our TEC protocol. The error detection and correction circuit is demonstrated in the D and E block of the circuit. The quantum conditional operators for teleportation are demonstrated in block F of Fig. \ref{qtec_Fig2}. The conditional operators for Z-basis are implemented using $CNOT_{3,6}$, $CNOT_{4,7}$, $CNOT_{5,8}$ gates and for X-basis are implemented using $H_1$, $H_2$, $H_3$, $H_6$, $CNOT_{0,6}$, $H_7$, $CNOT_{1,7}$, $H_8$, $CNOT_{2,8}$, gates respectively. See the Results Section \ref{qtec_Results} for more details.

\textbf{TEC of erasure errors}
We encode a data qubit as given in Eq.\eqref{qtec_Eq3}. The encoding is demonstrated in block A of Fig. \ref{qtec_Fig3}. The information is encoded for $\alpha=0.92$ and $\beta=0.38$, like we did in the case of TEC circuit for bit-flip errors using the gates $T_0$, $H_0$, $S_0$ in succession. The $H_0$ gate is used to create a superposition of $(\Ket{0}+\Ket{1})/\sqrt{2}$ from the initial $\Ket{0}$ state. The gates $CNOT_{0,1}$, $CNOT_{0,2}$, $CNOT_{0,3}$, $CNOT_{0,1}$, $CNOT_{2,3}$, $H_0$, $H_2$, $CNOT_{0,1}$ and $CNOT_{2,3}$ are used to encode a data qubit $(\Ket{0}+\Ket{1})/\sqrt{2}$ into a logical qubit of the form $\frac{\alpha}{2} (\ket{00}+\Ket{11})_{12}(\ket{00}+\Ket{11})_{34} + \frac{\beta}{2} (\ket{00}-\Ket{11})_{12}(\ket{00}-\Ket{11})_{34}$. This circuit is used for encoding our information in the form of physical qubits. The logical Bell state is encoded using gates, $H_4$, $H_{11}$, $CNOT_{4,5}$, $CNOT_{4,6}$, $CNOT_{11,10}$, $CNOT_{11,9}$, $CNOT_{6,7}$, $CNOT_{9,8}$, $H_9$, $X_9$, $CNOT_{9,7}$ and $CNOT_{9,6}$ gates as shown in block B of Fig. \ref{qtec_Fig3}. The QND measurement for detecting erasure errors are demonstrated in block C of Fig. \ref{qtec_Fig3} by using the gates $CNOT_{1,12}$, $CNOT_{2,13}$, $CNOT_{3,14}$ and $CNOT_{3,15}$. Since the number of qubits are limited to 14, we didn't implemented these gates on the real device during execution. The logical CNOT gates are implemented using $CNOT_{0,4}$, $CNOT_{1,5}$, $CNOT_{2,6}$ and $CNOT_{3,7}$ as illustrated in block D of Fig. \ref{qtec_Fig3}. The quantum conditional operators for teleportation are demonstrated in block E of Fig. \ref{qtec_Fig3}. The conditional operators for Z-basis are implemented using $CNOT_{5,9}$, $CNOT_{6,10}$, $CNOT_{7,11}$ gates and for X-basis are implemented using $H_1$, $H_2$, $H_3$, $H_9$, $CNOT_{1,9}$, $H_{10}$, $CNOT_{2,10}$, $H_{11}$, $CNOT_{3,11}$ gates respectively. See the Results Section \ref{qtec_Results} for more details.

\section*{Data availability}
Data are available to any reader upon reasonable request.

\section*{Acknowledgments\label{qtec_Acknowledgments}}
K.M.A. and M.S. would like to thank Indian Institute of Science Education and Research Kolkata for providing hospitality during the course of the project. B.K.B. acknowledges the support of IISER-K Institute Fellowship. The authors acknowledge fruitful discussions with Dr. Sreraman Muralidharan for the improvement of the manuscript. The authors acknowledge the support of IBM Quantum Experience for producing experimental results. The views expressed are those of the authors and do not reflect the official policy or position of IBM or the IBM Quantum Experience team.

\section*{Author contributions}
K.M.A. has developed the quantum error correction codes. K.M.A. and M.S. have discussed and designed all the quantum circuits, K.M.A. has analysed the data. K.M.A. has performed all the experiments in IBM Quantum Experience platform. M.S. has drawn all the quantum circuits using sharelatex codes. K.M.A., M.S. and B.K.B. have the contribution to the composition of the manuscript. B.K.B. have supervised the project. P.K.P. has thoroughly checked and reviewed the manuscript. K.M.A, M.S., and B.K.B. have completed the project under the guidance of P.K.P.  

\section*{Competing interests}
The authors declare no competing financial as well as non-financial interests. 

\iffalse
\section{Supplementary Information: Demonstration of a general fault-tolerant quantum error detection code for $(2n+1)$-qubit entangled state on IBM 14-qubit quantum computer}
For simulating the error detection protocol, we used QISKit to take both simulation results. The QASM code for the same is as follows: 

\lstinputlisting[language=Python]{errordetection.py}
\subsection{Measurement data}
We performed all the simulations on QISKit and recorded the countings of each of the measurement result over the two ancillary error syndrome qubit in 8192 shots. From the countings, the probability of each error \textit{i.e.} bit-flip error, phase-flip error and arbitrary phase-change error was extracted. The data is shown in the table \ref{qed_table4} below.
\fi
\end{document}